\documentclass{sf2a-conf2018}
\usepackage{graphicx}
\usepackage{hyperref}
\usepackage[]{natbib}  
\usepackage{epstopdf}

\def\BibTeX{{\rm B\kern-.05em{\sc i\kern-.025em b}\kern-.08em
    T\kern-.1667em\lower.7ex\hbox{E}\kern-.125emX}}
\bibpunct{(}{)}{;}{a}{}{,}  
\newcommand{\kms}{{\mathrm{km\cdot s^{-1}}}}

\newcommand{\vsini}{\ensuremath{v_{{\mathrm e}}\cdot\sin i}}
\newcommand{\teff}{\ensuremath{T_\mathrm{eff}}}
\newcommand{\logg}{\ensuremath{\mathrm{log}\ g}}

\newcommand{\logXH}[1]{log(\frac{n_{#1}}{n_H})}
\newcommand{\logbck}[1]{ {[} \frac{n_{#1}}{n_H} {]} }

\DeclareRobustCommand{\ion}[2]{\textup{#1\,\textsc{\lowercase{#2}}}}

\begin{document}

\TitreGlobal{SF2A 2018}


\title{Line synthesis of several chemical elements from Carbon to Bismuth in the spectrum of HD72660}

\runningtitle{Determination of the abundances of several chemical elements of HD72660}

\author{A. Lemonnier}\address{LESIA, UMR 8109, Observatoire de Paris Meudon, Place J.Janssen, Meudon, France}

\author{R. Monier}\address{LESIA, UMR 8109, Observatoire de Paris Meudon, Place J.Janssen, Meudon, France}



\setcounter{page}{237}

\maketitle


\begin{abstract}
The high resolution and high $S/N$ HARPS and STIS spectra of the A0Vm star HD72660 have been synthesised using model atmospheres and the line synthesis code Synspec49 in order to derive the abundances of several key chemical elements. 
In particular we have derived abundances of elements which have their strongest lines in the UV like Ru, Yb, Au, Pb and Bi. 
We find overabundances of heavy elements in particular strong overabundances of Pt, Pb and Bi which are probably the signature of radiative diffusion in this star.
\end{abstract}

\begin{keywords}
stellar atmosphere, abundances, stars: individual: HD72660, stars: chemically peculiar
\end{keywords}

\section{Introduction}
Previous studies of the hot A0Vm star HD72660 have reported overabundances of iron-peak elements and pronounced overabundances of heavy elements in its atmosphere , see  \cite{Monier2018}.
The low rotational velocity of this star favors a radiative atmosphere little mixed by rotation. 
Overabundances and underabundances probably reflect an efficient action of radiative acceleration on these heavy elements which have rich transitions, accumulating these elements in the line forming region.\\
The aim of this work is therefore to provide determinations of new abundances of heavy elements.

\section{Observations and reduction}
The observed UV spectra of HD72660 were obtained by Ruth Peterson and has been retrieved from the MAST archive. 
They were recorded with STIS at a high resolution ($R=114000$). 
The UV range enables to study ions which hardly have  any lines in the optical range.
As a starting mixture of abundances we have used the abundances derived from the optical HARPS spectra ($R=115000$) by \cite{Monier2018}.   
The $S/N$ of STIS and HARPS spectra are respectively  $S/N \simeq 120$ and $S/N \simeq 146$, appropriate for abundance analyses.

\section{Synthetic spectrum computations and abundance determinations}
The fundamental parameters have been derived in \cite{Monier2018}: $\teff=9650\pm250$ K, $\vsini=5.0\pm0.5\,\kms$, $\logg=4.05\pm0.25$ dex and $\xi=2.20\pm0.20\,\kms$.\\
We computed a model atmosphere with \citet{Kurucz92} ATLAS9 code with 72 parallel layers assuming Local Thermodynamical Equilibrium (LTE), Radiative Equilibrium (RE) and Hydrostatic Equilibrium (HE). 
Synthetic spectra were computed using \citet{Hubeny95} Synspec49 code by using as first solution the abundances produced from the synthesis of the HARPS spectrum of HD 72660 in \citet{Monier2018}.\\

Saha's ionisation equation has been used to compute the ionisation ratios with depth.
We have assumed that ionisation energies $\chi$ are constant with temperature. The partition functions $Z$ of ions were computed by fitting polynomials. For bismuth, we find that \ion{Bi}{I} is dominant near the surface and \ion{Bi}{II} dominates deep in the atmosphere where collisions are more important than radiative processes and therefore LTE prevails.  We find that \ion{Bi}{III} is negligible throughout the entire atmosphere. It is therefore justified to choose \ion{Bi}{II} lines to derive the Bi abundance.\\
The \ion{Bi}{II} line at $\lambda=1791.84$ \AA\ was used to derive the bismuth abundance. A pair of \ion{Fe}{II} lines on the blue and red side of the synthesised line was used as control lines to establish an accurate wavelength scale.
The synthesis of this line reveals a large overabundance of Bismuth: $\logXH{Bi}=-9.51\pm0.20$ dex. An uncertainty of about $\pm0.20$ dex was computed with an arithmetic mean weighted by uncertainty on each line when available.
This line is blended with a few weak lines of Fe, Cr and Mn, which contribute little to the opacity. \\


For all elements, the abundance excesses  $\logbck{X}$\footnote{$\logbck{X}:=\logXH{X}-\logXH{X}_\odot$} have been computed with respect to the solar abundances of \citet{Grevesse98}.
We find that carbon is underabundant $\logbck{C}=-1.05$ dex. We also find that heavy elements have significant overabundances as expected for Am stars. 
Zirconium is overabundant by a factor 25 times the solar abundance (from the line at 1790.113 \AA), ruthenium is overabundant also by a factor 25  (from the line at 1883.06 \AA), ytterbium by a factor 20 (from the line at 1873.879 \AA),  platinum by a factor 71 (from the line at at 1873.879 \AA), gold by a factor 89 (from the line at 1740.47 \AA) and lead by a factor 71 (from the line at 1682.12 \AA). These abundances are compared to the solar abundances and collected in Table 1.


\begin{table}
\centering
\begin{tabular}{|l|l|l|p{3cm}|}
 \hline
  \multicolumn{3}{|c|}{Determined abundances in HD 72660}   \\ \hline
  \multicolumn{1}{|c|}{Element} &
  \multicolumn{1}{|c|}{Abundance} &
  \multicolumn{1}{|c|}{Solar abundance} \\
 \hline
C   & -4.48 & -3.48   \\
Zr  & -7.29  & -9.40     \\
Ru  & -8.76 & -10.76   \\
Yb  & -9.62 & -10.92  \\
Pt  & -8.90  & -10.20      \\
Au  & -9.04  & -10.99      \\
Pb  & -9.05  & -10.91    \\
Bi  & -9.51  & -11.29   \\  
\hline
\end{tabular}
\caption{Abundance analysis for HD 72660}  
\end{table}

\section{Conclusions}
New abundances for several elements which have few lines (Zr, Ru, Yb, Pt, Au, Pb and Bi) in the optical range have been derived.
The found overabundances for the very heavy elements (and the underabundances for light elements) suggest an efficient action of radiative diffusion which 
support these elements in the line formation region of HD72660.
Future line synthesis of the UV and optical spectrum of HD72660 is envisaged to derive more abundances from this very rich spectrum.

\begin{acknowledgements}
The ESO archive has been queried for the HARPS spectrum of HD72660. MAST\footnote{\url{https://mast.stsci.edu/portal/Mashup/Clients/Mast/Portal.html}} was used to fetch UV spectra of HD72660.
The authors have used the NIST \footnote{\url{https://www.nist.gov/pml/atomic-spectra-database}} Database and the VALD3\footnote{\url{http://vald.astro.uu.se}} database operated at Uppsala University to upgrade atomic data.
\end{acknowledgements}


\end{document}